\documentclass{article}
\usepackage{url}
\usepackage{amsmath}
\usepackage{graphicx}
\begin{document}
%\documentclass{article}
%\usepackage{url}
%\usepackage{amsmath}
%\usepackage{graphicx}
%\pagenumbering{arabic}
%\thispagestyle{empty}
%\begin{document}
%\vspace{2.0ex}
%Changed 30th January 2017
%
%\begin{document}
\vspace*{-3cm}
\begin{flushright}
\today
\end{flushright}
\vspace{3.0ex}
\begin{center}  
\begin{Large}
{\bf PHYSTAT$\nu$ at CERN (January 2019)\footnote{The NuPhys Workshop
took place before the PHYSTAT$\nu$ meeting, but this article was written
after it took place, and so reports on the actual meeting.} }
\end{Large}
\end{center}
%\vspace{0.5in}
\begin{center}
\vspace{2.0ex}
\vspace{1.0ex} {\Large Louis Lyons}\\
Imperial College, London and Particle Physics, Oxford\\
\vspace{5.0ex}
%{\Large ABSTRACT}
ABSTRACT 
\end{center}
A short overview is provided of the recent PHYSTAT$\nu$ meeting at CERN,
which dealt with statistical issues relevant for neutrino experiments.

%\section{Introduction}
%PHYSTAT$\nu$ after NuPhys. This note written after, so reports on meeting.
\section{PHYSTAT}
The PHYSTAT Workshops\cite{CERN_CLW} - \cite{PHYSTATnuCERN} have concentrated on the {\bf statistical} issues 
that arise in Particle Physics
analyses, rather than on the experimental {\bf results}. They started in 2000, with meetings on 
`Confidence Limits' at CERN\cite{CERN_CLW} and at Fermilab\cite{FNAL_CLW}. More recent meetings have 
concentrated on specific types of 
experiments, with several being on Collider analyses. There were also meetings in 2016 in Japan\cite{PHYSTATnuIPMU} and at
Fermilab\cite{PHYSTATnuFNAL} on statistical issues relevant for neutrino experiments. (Another is planned for Dark 
Matter experiments in summer 2019 \cite{PHYSTAT-DM}.)

Since CERN has played a prominent role in neutrino experiments, it was felt very appropriate to have a third 
PHYSTAT$\nu$, at CERN\cite{PHYSTATnuCERN}.  The meeting was from $23^{rd}$ to $25^{th}$ January 2019.

\section{Structure of PHYSTAT$\nu$ at CERN}
Apart from invited and contributed talks and a poster session, there were several different components that made up 
the meeting:
\begin{itemize}
\item{Before the main part of the Workshop started, there were two well-attended talks on basic statistics, for 
those who wanted a reminder of some of the relevant statistical issues. }
\item{ The Workshop started with two introductory talks. The first was about  interesting historical
and contemporary neutrino experiments, while the second covered the range of statistical issues relevant for 
neutrino analyses to be discussed at the Workshop.}
\item{There were three quarter-day sessions devoted to specific topics: Unfolding, Systematics and Machine 
Learning (see below). Each of these had an overview talk, followed by the Collider experience and then
practical applications in neutrino experiments.} 
\item{As at  other PHYSTAT meetings, the participation of Statisticians added greatly to the value of
the meeting. We are very grateful to Jim Berger, Anthony Davison, Michael Kuusela, Victor
Panaretos, Chad Schafer and David van Dyk for being at the Workshop.  Not only did we benefit from 
their relevant talks, but is was also extremely useful to have them 
available for informal discussions between the sessions. }
\item{At the end of each of the first two days, Tom Junk reviewed the day's highlights, and stimulated discussion 
on them by the audience. His insights were very useful.}
\item{The Workshop ended with two summary talks, by Statistician van Dyk and by Physicist Kevin MacFarland.
They reminded and entertained us of the Workshop's events, and gave us thoughts to consider for the future.}
\end{itemize}
The slides and videos of the talks are available at the Workshop's 
website \url{https://indico.cern.ch/event/735431/timetable/}

\section{Neutrino statistical issues}
\subsection{General} 
A few topics recurred throughout the Workshop.
\begin{itemize}
\item{Combining different measurements: Sometimes there is more than one measurement of a physical quantity that is required 
for a current analysis e.g. old measurements of cross-sections of neutrinos on various nuclei. A problem arises 
in how to deal with discrepancies among separate measurements. An ad hoc approach is to expand the uncertainties on the combined result to compensate for the discrepancies.
The problem is aggravated by absence of information 
about correlations among the different data measurements in a single experiment, or even between different experiments. 
There is no real statistics solution to this. It is not necessarily conservative to set correlation coefficients to 
zero (or to unity).} 
\item{$5\sigma$ for discovery? Reasons suggested for this stringent criterion for claiming a discovery include (a) past false claims of discovery; (b) the `Look Elsewhere Effect'; (c) underestimated systematics; (d) the idea that `Extraordinary discovery 
claims require extraordinary evidence'; etc.\cite{LL5sigma}  Even though not all experiments are equally affected by these features,
it is hard to see universal agreement on a different significance level being adopted for different types of measurements.
Nevertheless it seems unreasonable to demand evidence at the $5\sigma$ level for the discovery of CP-violation, or
for choosing between the normal and inverted mass hierachies for neutrinos.} 

\item{Asimov or toys for expected results: It is common to quote not only an interval for a measured parameter (or an upper
limit)  using the actual data, but also the expected sensitivity of the experiment. This is some sort of representative 
value summarising the range of results that might be obtained if the measurement were to be repeated many times. This is
useful (i) for comparing with the corresponding observed quantity from the actual data, to check that the latter is not unreasonable; and (ii) as a means of funding bodies comparing the expected results from competing proposed experiments. 
 
Often the median result from many simulations is used to quantify the sensitivity. An alternative that requires less computation 
is to use `Asimov data'. This is a single data set, in which statistical fluctuations have been eliminated. i.e. If the 
expected number of events in a bin is 6.3, the Asimov data set contains 6.3 events there. Hopefully Asimov data and 
Monte Carlo toys provide similar estimates of the sensitivity, but care is necessary as this is not always the case. }

\item{$p$-values and likelihood ratios: In searches for new physics, it is often noted that $p$-values tend to be smaller than likelihood ratios. Sometimes there is an underlying implication that this shows that $p$-values tend to over-emphasize the evidence in favour of a new discovery.  In fact there is absolutely no reason why they should agree:
 the $p$-value refers to the possible disagreement of the data with just the null hypothesis (i.e. just conventional
physics), while the likelihood ratio compares how well the two hypotheses (conventional physics versus something new) explain the data. Indeed
for a given value of $p$, the likelihood ratio can take on a range of values, depending on the separation of the probablity density
functions of the data statistic for the two hypotheses. It is rather like trying to decide whether the number of protons in an elephant or the ratio 
of its height to that of a mouse is better for assessing its size.}

\item{Neutrino masses: These can be derived from the difference in mass-squared of the various neutrino pairs, as measured from neutrino oscillation data; and cosmological information on the sum of the neutrino masses. The latter is estimated, albeit somewhat controversially, as being less than 120 milli-eV/$c^2$. Assuming that this is true, the masses can be determined with reasonable precision. Bayesian methods have been used for this, with inevitably discussion on the dependence of the results on the 
priors (see, for example, ref. \cite{nu_masses}).
However,  a simple back-of-envelope calculation with no priors involved can be performed: the mass of the lightest neutrino can range from zero up to a value such that the known mass-squared differences then produce masses which saturate the comological limit on their sum. This gives results which are very much in line with the Bayesian estimates. However, the choice of prior is more important for comparing whether data favour the normal or the inverted ordering of neutrino masses. This is in line with the fact that the choice of prior is more important for Hypothesis Testing than for Parameter Determination.}
    
\end{itemize}

\subsection{Unfolding}
Experimental detectors do not provide perfect measurements of the  objects produced in the observed interactions.
One procedure to correct for this is to use an unfolding technique, involving the known experimental
resolution. This is often a tricky process, and estimating the uncertainties on an unfolded result can be problematic.
Furthermore, even though the number of entries in the different bins of a histogram may be uncorrelated, this is 
not so for the unfolded spectrum. Nevertheless, the unfolded spectra should be better estimates of the `true 
distributions of Nature' than the original spectra, and so are considered useful as information to be transmitted to posterity.

However, in comparing data with some theoretical prediction, it is easier to {\bf smear} the theory, than to 
{\bf unfold} the data. This can even be done with future theories, provided the experiments provide information 
about their smearing matrices.   

Another issue is the choice of bin-width for the unfolded spectrum. If they are too narrow, there are big smearing effects from one bin to its neighbours, so that unfolding has a big effect. If on the other hand they are too wide, one loses the opportunity to observe fine structure in the unfolded spectrum; and the smearing matrices become model dependent, as they depend on the actual distribution of data within each bin.

The situations in which it seems that unfolding is necessary are for comparing two experiments measuring the same
physical quentity, but with different resoltions; and using the data to help tune Monte Carlo simulations, where re-smearing at each iteration of the optimisation may be computationally too expensive. 
  
The conclusion is that it is desirable  to make available in publications the original data,  its unfolded version
with its covariance matrix, and the smearing matrices. 

\subsection{Systematics}
Neutrino experiments range from having very high statistics (e.g. the near detectors at accelerators and at reactors) to those with very few (or perhaps even zero!) signal events, e.g. searches for neutrino-less double beta decay, or for neutrinos from supernovae. The former tend to be dominated by systematics effects, while for the latter statistical fluctuations are likely to be more important.

%Range of experiments in distance from source to detector, and size of data set
In dealing with systematics, it is usual to employ Bayesian techniques, as they tend to deal better with large numbers of parameters. Even when a frequentist approach is used for the parameter(s) of interest, often a Bayesian method is used to eliminate the
systematic effects' nuisance parameters. Such an approach was orginally suggested by Cousins and Highland in the context of setting upper limits\cite{Cousins-H}. 

The elimination of the nuisance parameters is achieved via profiling or marginalisation. The former is used for likelihoods, and  for each value of the physics parameter $\phi$ it involves calculating the likelihood for the best value of the nuisance parameter $\nu$. i.e.
\begin{equation}
L_{prof}(\phi) = L(\phi, \nu_{best}(\phi), 
\end{equation}
where the best value of $\nu$ can be a function of $\phi$. In contrast, marginalisation is for posterior probabilities $P(\phi, \nu)$,
which are integrated over $\nu$ to give $P_{marg}(\phi)$.

%Uncertainty on uncertainty

\subsection{Machine Learning}
There has recently been a rapid increase in the popularity and use of deep neural network machine learning 
techniques (see, for example, the lecture series at CERN and at Fermilab\cite{CERN_ML, FNAL_ML}. 
In Particle Physics they have been used for on-line triggering, tracking, fast simulation, 
object identification, image recognition (e.g. for an interaction in a large liquid Cherenkov detector, how many rings are there 
and what particles caused them?) and event-by-event separation of signal from background. Because they are 
potentially so powerful, it is important to have a set of protocols to check that they perform in a reliable manner,
and are not introducing subtle biases of which the user is unaware.

For neural networks with a single hidden layer, it is easy to understand how they are capable of selecting a 
particular region of the multi-dimensional space for the input variables that are enhanced in signal (with 
respect to background). This helps to some extent in choosing the particular architecture of the net to use. 
Unfortunately this is very difficult to achieve for deep networks. 
 
\section{Conclusions}
\begin{itemize}
\item{The participation in PHYSTAT$\nu$ of several physicists working at Hadron Colliders was very useful, 
as many of the statistical issues are common to the two fields. }

\item{Hadron Collider experiments, each with thousands of physicists, can afford a few to spend part of their 
time in the collaboration's statistics committee. Neutrino experiments are probably too small for each to have 
its own statistics committee, but interest was shown in their being such a committee for the neutrino 
community as a whole. Worries were expressed about the individual experiments' confidentiality requirements, but 
it is usually possible to discuss a statistical issue without revealing too many physics secrets. Also it is 
becoming common for different experiments to perform a combined analysis\footnote{This means performing a 
physics analysis on the combined {\bf data}, rather the inferior procedure of trying to combine the 
{\bf results.}} on some specific topic, and having a 
joint statistics committee is a good start to such a process.}

\item{Most of the participants were happy with the Workshop and found it useful. There seemed to be general support for having another PHYSTAT$\nu$, perhaps in two or three years time.}
%Square wheels
\end{itemize}

%%%%%%%%%%%%%%%%%%%%%%%%%%%%%%%%%
\begin{figure}[ht]
\begin{center}
\includegraphics[width=\textwidth]{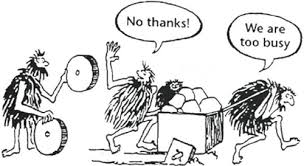}
\caption{Illustration from Kevin McFarland's summary lecture of a comment in the basic statistics introduction that, in the choice of a statistical technique, ``it is better to use a Statistician's round wheel than your own square one".  }
\label{SquareWheels}
\end{center}
\end{figure}
%%%%%%%%%%%%%%%%%%%%%%%%%%%%%%%%%%%%%

%%%%%%%%%%%%%%%%%%%%%%%%%%%%%%%%%%%%%
\end{document}